\documentclass[journal]{IEEEtran}
\IEEEoverridecommandlockouts 

\usepackage{lipsum}
\usepackage{etoolbox}

\usepackage{cite}

\usepackage{hyperref}

\usepackage {siunitx}

\usepackage{float}

\usepackage[inline]{enumitem}

\ifCLASSINFOpdf
   \usepackage[pdftex]{graphicx}
   \graphicspath{{./Images/}}
   \DeclareGraphicsExtensions{.pdf,.jpeg,.png}
\else
\fi
\usepackage[cmex10]{amsmath}
\usepackage{booktabs,colortbl}

\usepackage{amsmath}
\usepackage{amssymb}
\usepackage{stackengine}
\newcommand\barbelow[1]{\stackunder[1.2pt]{$#1$}{\rule{.8ex}{.075ex}}}
\usepackage{float}

\usepackage[version=3]{mhchem} 
\begin{document}

%
\title{Impacts of Community and Distributed Energy Storage Systems on Unbalanced Low Voltage Networks}
\author{\IEEEauthorblockN{Yiju Ma, Mohammad Seydali Seyf Abad, Donald Azuatalam, Gregor Verbi\v{c}, Archie Chapman}\\
	\IEEEauthorblockA{School of Electrical and Information Engineering, The University of Sydney, Sydney, Australia\\}
	\IEEEauthorblockA{Emails:\{yiju.ma, mohammad.seidaliseifabad, donald.azuatalam, gregor.verbic, archie.chapman\}@sydney.edu.au\\}}

%


\maketitle

\begin{abstract}
	Energy storage systems (EES) are expected to be an indepensable resource for mitigating the effects on networks of high penetrations of distributed generation in the near future. This paper analyzes the benefits of EES in unbalanced low voltage (LV) networks regarding three aspects, namely, power losses, the hosting capacity and network unbalance. For doing so, a mixed integer quadratic programmming model (MIQP) is developed to minimize annual energy losses and determine the sizing and placement of ESS, while satisfying voltage constraints. A real unbalanced LV UK grid is adopted to examine the effects of ESS under two scenarios: the installation of one community ESS (CESS) and multiple distributed ESSs (DESSs). The results illustrate that both scenarios present high performance in accomplishing the above tasks, while DESSs, with the same aggregated size, are slightly better. This margin is expected to be amplified as the aggregated size of DESSs increases. 
\end{abstract}


%

\section*{Nomenclature}
\addcontentsline{toc}{section}{Nomenclature}
\begin{IEEEdescription}[\IEEEusemathlabelsep\IEEEsetlabelwidth{$V_1,V_2,1$}]
	\item[$\mathcal{B}$] Set of total network buses 
	\item[$\mathcal{E}$] Set of potential network buses for an ESS installation		
	\item[$\mathcal{L}$] Set of total network lines 	
	\item[$\mathcal{T}$] Set of hours in one day 
	\item[$\mathcal{D}$] Set of days in one year	
	\item[$\mathcal{F}$] Set of network phases A, B and C
	\item[${f}$] Single network phase		
	\item[$p^\mathrm{load}_{t,i}$] Active load at time ${t}$, bus ${i}$  
	\item[$q^\mathrm{load}_{t,i}$] Reactive load at time ${t}$, bus ${i}$ 
	\item[$p^\mathrm{pv}_{t,i}$] Active PV generation at time ${t}$, bus ${i}$  
	\item[$q^\mathrm{pv}_{t,i}$] Reactive PV generation at time ${t}$, bus ${i}$ 
	\item[$p_{f,t,ij}$] Single phase active power at time ${t}$ between buses ${i}$ and ${j}$ 
	\item[$q_{f,t,ij}$] Single phase reactive power at time ${t}$ between buses ${i}$ and ${j}$ 	
	\item[$p^\mathrm{loss}_{t,ij}$] Total active power losses for all three phases at time ${t}$, between buses ${i}$ and ${j}$ 		
	\item[$p^\mathrm{loss}_{f,t,ij}$] Single phase active power loss at time ${t}$, between buses ${i}$ and ${j}$  
	\item[$v_{f,t,i}$] Phase voltage at time ${t}$, bus ${i}$ 	
	\item[$\bar{v}$] Maximum phase voltage for all buses 
	\item[$\barbelow{v}$] Minimum phase voltage for all buses 
	\item[$v_\mathrm{sub}$] Substation secondary winding phase voltage
	\item[$v^{+}_{t,i}$] Positive sequence voltage at time t, bus i	
	\item[$v^{-}_{t,i}$] Negative sequence voltage at time t, bus i
	\item[$R_{f,ij}$] Single phase resistance between buses ${i}$ and ${j}$   	
	\item[$X_{f,ij}$] Single phase reactance between buses ${i}$ and ${j}$ 	
	\item[$\Delta t$] Time slice			
	\item[$e_{t,i}$] State of charge at time ${t}$, bus ${i}$ 	
	\item[$\bar{e}$] Maximum state of charge 		
	\item[$\barbelow{e}$] Minimum state of charge  		
	\item[$e_\mathrm{0}$] Initial state of charge  	
	\item[$e_\mathrm{\Omega}$] Initial state of charge for the next cycle  	
	\item[$\eta^{+}$] Energy storage charging efficiency  	
	\item[$\eta^{-}$] Energy storage discharging efficiency  
	\item[$p^{+}_{t,i}$] Charging rate for one phase at time ${t}$, bus ${i}$ 	
	\item[$p^{-}_{t,i}$] Discharging rate for on phase at time ${t}$, bus ${i}$ 		
	\item[$\bar{p}^{+}$] Maximum charging rate 	
	\item[$\bar{p}^{-}$] Maximum discharging rate 
	\item[$e^\mathrm{dist}_{i}$] Capacity for the ${i^{th}}$ distributed energy storage system
	\item[$e^\mathrm{com}$] Capacity for a community energy storage system  		 		
	\item[$B^\mathrm{loc}_{t,i}$] Binary variable to locate an energy storage at bus ${i}$ 
	\item[$N$] Number of storage systems on the network		
	\item[$\mathrm{VUF}$] Voltage unbalance factor 
\end{IEEEdescription}

\section{Introduction}
Low carbon technologies (LCTs) such as photovoltaic (PV) systems and wind generation have been widely employed as the use of conventional power sources continues to be eliminated. However, high residential PV generation can lead to a variety of negative impacts such as over-voltage, phase unbalance and power quality issues on LV networks \cite{tonkoski2012impact}, \cite{watson2016impact}.

ESS are considered as a useful technology as it can benefit distribution networks in many aspects. For example, they can maximize the utilization of intermittent renewable generation, reduce peak loading and remove line congestions. These together can lead to power loss reduction and network upgrade deferral, thereby improving the efficiency with which electricity is delivered to customers \cite{chu2012opportunities}. ESS can also provide enhancement to grid flexibility and ensure a higher security of energy supply during undesired weather conditions \cite{atwa2010optimal}. More importantly, they can mitigate over voltages by absorbing excessive renewable generation in LV networks, and release the stored energy whenever it is needed \cite{eyer2010energy}. Considering these benefits, the value of ESS has risen significantly, especially in Australia, where PV penetration is significantly higher than the rest of the world \cite{condon2016customer}. However, there is still considerable debate as to whether a CESS or DESSs is more valuable to our networks, and currently few studies have compared their benefits and drawbacks, as reviewed below. 

Numerous studies have presented optimisation models and have shown their potential in determining the optimal sizing and placement of ESS under different circumstances. Authors in \cite{pandvzic2015near} use a DC optimal power flow (DC OPF) model for deciding the optimal ESS sizing and placement. This model is based on the linearization of AC optimal power flow (AC OPF) and it presents limited computational complexity. However, this model is only suitable for transmission networks. A full AC OPF can assist in achieving accurate results for distribution networks, but the problem often becomes intractable due to its high complexity. This issue can be overcome by adopting convex relaxation, suggested in \cite{nick2014optimal} and \cite{low2014convex}. Relaxations are widely used for linearizing higher-order terms in objective functions and constraints, which often result in a second order cone model. The study in \cite{nick2014optimal} employs it to illustrate the effectiveness of ESS in voltage regulation, congestion elimination and power loss reduction. However, this method is not applied to LV networks. A decoupled AC OPF incorporated with genetic algorithm (GA) is applied to determine the optimal sizing of ESS in \cite{abad2016optimal}, with the objective being minimizing the electricity provision cost for network operators. The drawbacks of this paper include:
\begin {enumerate*} [label=\itshape\alph*\upshape)]
\item[(i)] the study is conducted on an MV network, \item[(ii)] the problem can become difficult to solve if a complex network is involved, \item[(iii)] a heuristic method (GA) is used.
\end {enumerate*} 

The study in \cite{poulios2015optimal} describes a linearized AC power flow model for optimal ESS sizing and location on a LV feeder based on cost minimization. It demonstrated the great potential of ESS in boosting the hosting capacity with the aid from active power curtailment (APC). However, this paper does not consider power loss minimization, and the network is assumed to be balanced.      

Many studies do not consider optimal sizing and location, and instead, they propose a range of heuristic control strategies to maximize the benefits of ESS. For instance, \cite{zeraati2016distributed} illustrated that ESS with the proposed coordinated control strategy can effectively mitigate over votlages when exposed to high PV generation. Authors in \cite{marra2014decentralized} proposed a decentralized storage control strategy based on voltage sensitivity analysis in LV networks with high PV penetration. The results showed that high PV generation can lead to over sizing of ESS, which is uneconomical to operators. These studies do not optimally place and size the ESS, while the results can be significantly different otherwise.

Recently, \cite{fortenbacher2016optimal} studied the differences between a centralized ESS and multiple DESSs in a LV network. A forward-backward sweep linearised AC optimal power flow model was proposed for determining the optimal location and sizing for the DESSs. However, this study was conducted on a balanced network, and it failed to investigate the benefits of an optimally placed and sized centralized ESS. None of the papers above have explored the differences between a CESS and multiple DESSs when considering their impacts on power losses, hosting capacity and network unbalance as a whole.

In this paper, the impacts of ESS in power losses, the hosting capacity and network unbalance in LV networks are investigated. Specifically, two scenarios are examined:
\begin {enumerate*} [label=\itshape\alph*\upshape)]
\item[(i)] the installation of a single CESS, and \item[(ii)] the installation of multiple DESS with the same aggregated size.
\end {enumerate*} 
In order to determine the optimal location and the associated capacity of the ESS in both scenarios, a linearized approximation of AC OPF is applied. The optimization model aims to minimize annual energy losses while satisfying voltage constraints. Given the results of simulations, we assess, quantify and compare the effectivenesses of these two ESS configurations.

The outcomes of this research illustrate that the distributed configuration has an edge over the community configuration in accomplishing the tasks described above, although both of them are highly effective. However, DESSs present higher flexibility, specifically, they are expected to be more effective as the aggregated size increases.    

The rest of the paper is organised as follows. Section 2 describes the ESS modelling. Section 3 details the problem formulation and the methodology for determining the optimal location and the associated ESS capacity. Section 4 includes a detailed explanation of the network model and a discusstion on the simulation results. Section 5 draws conclusions.

\section{Energy Storage Modelling}
 
The ESS model is presented in this section. The state of charge at time ${t}$ of an ESS, placed at bus ${i}$, is a function of the state of charge at time ${t-1}$ and the charging and discharge rates during this time interval. The charging and discharging rates discovered are for one phase only, thus, they must be multiplied by 3: 

\vspace{-0.3cm}

\begin{equation} \label{eq1}
	e_{t,i} = e_{t-1,i} + 3\big(\eta^{+}p^{+}_{t,i} - \frac{1}{\eta^{-}}p^{-}_{t,i}\big)\Delta t.
\end{equation}

The initial state of charge after each cycle, typically one day, must be the same, this is formulated as follows:
\vspace{-0.1cm}
\begin{equation}\label{eq2}
	e_{\Omega} = e_\mathrm{0}.
\end{equation}
\vspace{-0.1cm}
To ensure that the above constraint can be satisfied, the total charged and discharged power for one cycle must be the same:

\begin{equation}\label{eq3}
	\sum\limits_{t \in \mathcal{T}} \Big(\eta^{+}p^{+}_{t,i} - \frac{1}{\eta^{-}}p^{-}_{t,i}\Big) = 0.
\end{equation}

In addition, charging and discharging rates at each time interval ${\Delta t}$ must be constrained the maximum charging and discharging rates stated by the properties of the ESS:   

\begin{equation}\label{eq4}
	0 \leq 3p^{+}_{t,i} \leq \bar{p}^{+}.
\end{equation}
\vspace{-0.2cm}
\begin{equation}\label{eq5}
	0 \leq 3p^{-}_{t,i} \leq \bar{p}^{-}.
\end{equation}

Similarly, the state of charge cannot exceed the minimum and maximum states of charge at all times, indicated by the equation below:    

\begin{equation}\label{eq6}
	\barbelow{e} \leq e_{t,i} \leq \bar{e}.
\end{equation}

In this study, the charging and discharging efficiencies in the model are neglected.

\section{Mathematical Fomulation}
The optimisation model aims to place and size a CESS and multiple DESSs in an unbalanced LV network in order to minimize annual system losses.  
\vspace{-0.1cm}
\subsection{Objective function}
The objective is to minimise system losses, formally stated as:
\begin{align}\label{eq7}
\underset{\substack{p^{+}_{t,j}, p^{-}_{t,j}}, B^{\mathrm{loc}}_{t,j}, e^{\mathrm{com}}, e^{\mathrm{dist}}_{j}}{\text{minimize}}  \sum\limits_{d \in \mathcal{D}}\sum\limits_{t \in \mathcal{T}} \sum\limits_{i,j \in \mathcal{L}} p^\mathrm{loss}_{t,ij},
\end{align}

where $p^{\mathrm{loss}}_{t,ij}$ is given by:

\begin{equation}\label{eq8}
	\begin{aligned}
	p^\mathrm{loss}_{t,ij} = \sum\limits_{f \in \mathcal{F}} R_{f,ij}\frac{p^2_{f,t,ij}+q^2_{f,t,ij}}{v^2_{sub}} 
	\quad\forall\ {f}\in\mathcal{F}.
	\end{aligned}
\end{equation}

\noindent This objective function indicates that the overall model is a quadratic problem.
\vspace{-0.1cm}
\subsection{Constraints}

\subsubsection{Branch flow equations}
Load flows in LV networks with numerous branches can be difficult to solve, especially when the network is unbalanced and each phase must be taken care of individually. In this case, the branch flow model introduced in \cite{farivar2012optimal} and \cite{baran1989optimal} is employed, which has been shown to be accurate in modelling power flows in radial distributed networks \cite{yeh2012adaptive}. The equations in this model show that power flows on each phase from bus ${i}$ to bus ${j}$, can be formulated as a function of power flows and power losses entering bus ${j}$, and the generation and consumption at bus ${j}$. The modified power flow equations with ESS placed on bus ${j}$, phase ${f \in \mathcal{F}}$ are presented below: 

\begin{equation}\label{eq10}
	\begin{split}
		\begin{aligned}
			p_{f,t,ij} &= \sum\limits_{j,k \in \mathcal{L}} \Big(p_{f,t,jk} + R_{f,ij}\frac{p^2_{f,t,ij}+q^2_{f,t,ij}}{v^2_\mathrm{sub}}\Big)\\ &\quad+ p^\mathrm{load}_{t,j} - p^\mathrm{pv}_{t,j} + \sum\limits_{j \in \mathcal{E}} (p^{+}_{t,j} - p^{-}_{t,j}),
		\end{aligned}
	\end{split}
\end{equation}

\begin{equation}\label{eq11}
	\begin{aligned}
		q_{f,t,ij} &= \sum\limits_{j,k \in \mathcal{L}} \Big(q_{f,t,jk} + X_{f,ij}\frac{p^2_{f,t,ij}+q^2_{f,t,ij}}{v^2_\mathrm{sub}}\Big)\\ &\quad+ q^\mathrm{load}_{t,j} - q^\mathrm{pv}_{t,j},
	\end{aligned}
\end{equation}

\begin{equation}\label{eq12}
	\begin{split}
		\begin{aligned}
		v^{2}_{f,t,j} &= v^{2}_{f,t,i} - 2(p_{f,t,ij}R_{f,ij} + q_{f,t,ij}X_{f,ij})\\ &\quad+ (R^{2}_{f,ij} + X^{2}_{f,ij})\frac{p^{2}_{f,t,ij} + q^{2}_{f,t,ij}}{v^{2}_{f,t,j}}.
		\end{aligned}
	\end{split}
\end{equation} 

In this paper, only the active power flow of ESS is considered. It has been clearly illustrated by multiple literatures that the nonlinear terms in equations \eqref{eq10} and \eqref{eq11}, representing power losses, can be neglected as the impacts of these terms on the objective function are limited \cite{yeh2012adaptive}. Under this asumption, the branch flow equations become:

\begin{equation}\label{eq13}
	\begin{split}
		\begin{aligned}
		p_{f,t,ij} &\approx \sum\limits_{j,k \in \mathcal{L}} p_{f,t,jk} + p^\mathrm{load}_{t,j} - p^\mathrm{pv}_{t,j} \\&\quad + \sum\limits_{j \in \mathcal{E}} (p^{+}_{t,j} - p^{-}_{t,j})
		\quad\forall\ {f \in \mathcal{F}}.\\
		\end{aligned}
	\end{split}
\end{equation}

\begin{equation}\label{eq14}
	\begin{aligned}
	q_{f,t,ij} \approx \sum\limits_{j,k \in \mathcal{L}} q_{f,t,jk} + q^\mathrm{load}_{t,j} - q^\mathrm{pv}_{t,j}
	\quad\forall\ {f \in \mathcal{F}}.\\
	\end{aligned}
\end{equation} 

\subsubsection{Voltage constraints}
Based on \cite{xu2017multi} and \cite{yeh2012adaptive}, equation \eqref{eq12} can be linearized by assuming voltage variations at an arbitrary bus ${j}$ are close to 0 at all times, eg. ${v_{f,t,j}-v_\mathrm{sub}\approx0}$. This assumption leads to the following expression: 

\begin{equation}\label{eq15}
	\begin{aligned}
	v^{2}_{f,t,j} \approx v^{2}_\mathrm{sub} + 2v_\mathrm{sub}(v_{f,t,j}-v_\mathrm{sub}) 
	\quad\forall\ {f \in \mathcal{F}}.\\
	\end{aligned}  
\end{equation} 

The linearized version of the voltage equation can be developed by substituting equation \eqref{eq15} to \eqref{eq12}:

\begin{equation}\label{eq16}
	\begin{aligned}
	v_{f,t,j} \approx v_{f,t,i} - \frac{p_{f,t,ij}R_{f,ij} + q_{f,t,ij}X_{f,ij}}{v_\mathrm{sub}}
	\quad\forall\ {f \in \mathcal{F}}.\\
	\end{aligned}
\end{equation} 

The accuracy of this equation has been justified by the authors in \cite{baran1989optimal}. In addtion, the model should satisfy the voltage limits for all three phases in presence of loads and PV generation:

\begin{equation}\label{eq17}
	\begin{aligned}
	\barbelow{v} \leq v_{f,t,i} \leq \bar{v}
	\quad\forall\ {f \in \mathcal{F}}.\\
	\end{aligned}
\end{equation}

\subsubsection{ESS location}
The following equation assists in deciding the optimal location for the CESS by setting the number of ESSs, ${N}$, to 1:

\begin{equation}\label{eq18}
	\sum\limits_{j \in \mathcal{E}} B^\mathrm{loc}_{t,j} = N.
\end{equation}

The ESS formulations from \eqref{eq1} to \eqref{eq6} and the branch flow equations \eqref{eq13}, \eqref{eq14}, \eqref{eq16}, along with votlage and ESS contraints \eqref{eq18} and \eqref{eq19} formulate a mixed integer quadratic programming model for determining the optimal location and the associated capacity for the CESS. 

The study aims to compare the impacts between a CESS and multiple DESS. Therefore, the next step is to determine the optimal locations and capacities for multiple DESSs, assuming they have the same aggregated capacity as the CESS. 

In this scenario, ${N}$ in equation \eqref{eq18} must be set as a variable instead of a parameter. Meanwhile, the following constraint which equates the aggregated capacity of the DESSs to the size of the CESS is added to the optimization model:

\begin{equation}\label{eq19}
	\sum\limits_{j \in \mathcal{E}} e^\mathrm{dist}_{j} = e^\mathrm{com}.
\end{equation} 

\section{Simulation and Results}

The modified MIQP with objective given by \eqref{eq7}, \eqref{eq8} and constraints \eqref{eq13}, \eqref{eq14}, \eqref{eq16}, \eqref{eq18}, \eqref{eq19} is modelled in AMPL\footnote{AMPL - Algebraic Mathematical Programming Language \cite{fourer1993ampl}}, a software which effectively models mathematical problems. Knitro\footnote{Knitro - A mixed integer non-linear solver within AMPL} is used to solve the optimization problem. Simulation is performed using Matlab, and the optimization model is accessed via AMPL API\footnote{API - Application Programming Interface}. All outcomes from both scenarios are discussed in this section.

\subsection{Test Network}
\begin{figure}[] 
	\includegraphics[width=8cm,keepaspectratio]{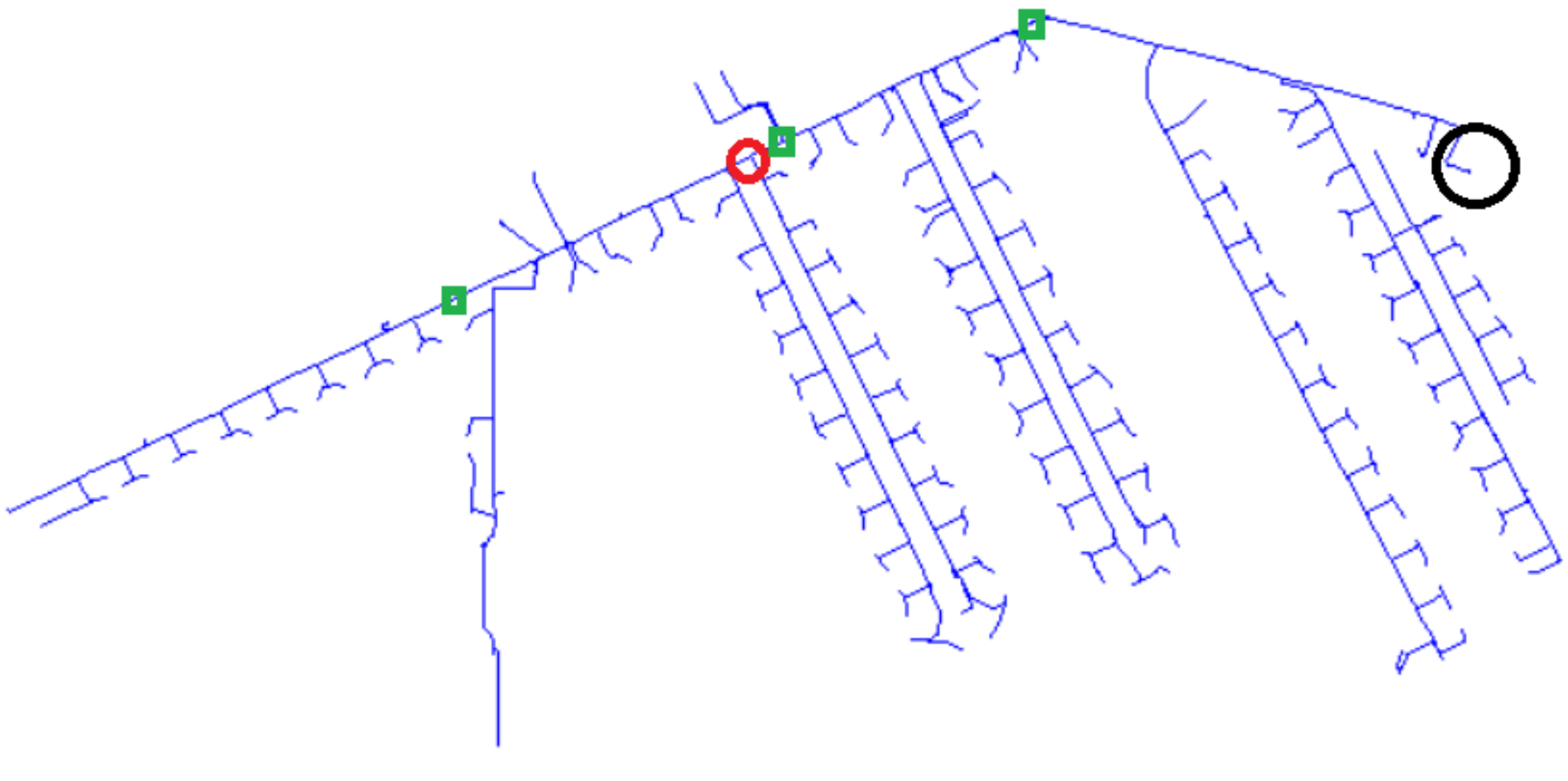}
	\caption{The unbalanced LV network, the black and red circles denote the substation transformer and CESS, while the three green squares represent DESSs \cite{enwlreport}}
	\label{fig1}
\end{figure}

The selected LV feeder is from the UK, which consists of ${175}$ single phase consumers with a total length of \SI{4.3}{\kilo\meter}. This is an unbalanced network with ${61}$, ${60}$ and ${54}$ consumers sitting on phase A, B and C, respectively. Each load, which follows a specific load profile, adopts a single phase PV system. This feeder covers a large area and it is reasonably loaded, presenting great potential for over-voltages. The network diagram obtained from OpenDSS\footnote{Open Distribution System Simulator \cite{opendss}} is shown in Fig.~\ref{fig1}.   

Load and PV profiles for one year are generated using the CREST model, which allows one to generate high resolution PV and load profiles \cite{richardson2010domestic}. For this study, each individual load and PV system are assigned with unique daily load and PV profiles with 1 hour resolution, respectively. 

The feeder data from OpenDSS are applied to AMPL when solving the optimization model. When performing simulations, the base voltage is \SI{0.4}{\kilo\volt} (\SI{1}{pu}) and the phase votlage of transformer secondary winding is fixed at \SI{0.23}{\kilo\volt}. 

\subsection{Performance Indices}
The three performance indices are:
\begin {enumerate*} [label=\itshape\alph*\upshape)]
\item power losses, \item the hosting capacity, and \item network unbalance.
\end {enumerate*} 

\subsubsection{Power Losses}
The primary task of the optimization model is to reduce annual network energy losses. For the base scenario, without any ESS, simulations are performed for a year. The total network losses in summer, autumn, winter and spring are approximatedly \num{2530}\si{\kilo\watt\hour}, \num{2820}\si{\kilo\watt\hour}, \num{4660}\si{\kilo\watt\hour} and \num{2190}\si{\kilo\watt\hour}, respectively. It can be seen that the total energy loss in winter is significantly higher than other seasons due to the employment of heating facilities. The annual network loss is \num{12200}\si{\kilo\watt\hour}, which is expected to be decreased by implementing the two ESS configurations.   

\subsubsection{Hosting Capacity}
The installed PV capacity is \num{175}\si{\kilo\volt\ampere} in the case that all PV system is rated at \si{1\kilo\watt}. In order to determine the hosting capacity, the rated power generation of PV systems on the entire feeder is gradually increased until the voltage threshold is met. This is most likely to occur at loads towards the end of the feeder. Simulation is performed in OpenDSS in which monitors are placed at all loads. An algorithm, which automatically increases PV generation until the voltage limit is exceeded, is implemented via the common interface shared between OpenDSS and Matlab. A typical summer day is tested using the algorithm since summer usually presents less network loading and reasonably high PV generation. The outcomes indicate that the upper limit is exceeded at bus ${2266}$ at 1pm when each rated PV generation is raised to \num{2.10}\si{\kilo\watt}. This means the \SI{100}{\%} penetration of this network without ESS is \SI{368}{\kilo\watt}.  

\subsubsection{Network Unbalance}
This study also investigates the effects of ESS on the voltage unbalance factor (${\mathrm{VUF}}$), which is a measure of the network unbalance \cite{pillay2001definitions}. It can be calculated through the following equation:
  
\begin{equation}\label{eq22}
\mathrm{VUF} = 100\frac{v^{-}_{t,i}}{v^{+}_{t,i}}.
\end{equation} 

The ${\mathrm{VUF}}$ at each bus is determined in OpenDSS. The maximum and average ${\mathrm{VUFs}}$ of the network without ESS are \num{0.401}\si{\%} at bus \SI{1940} and \num{0.275}\si{\%}, respectively. Altough the ${\mathrm{VUFs}}$ is currrently well below the UK standard of \si{2\%}, it can increase in the future as PV penetration increases.   
\vspace{-0.3cm}
\subsection{Community Energy Storage System}
The results of installing a CESS in the LV network are shown and analyzed in this subsection.

\subsubsection{Power Losses}
\begin{figure}[] 
	\centering
	\includegraphics[width=8cm,keepaspectratio]{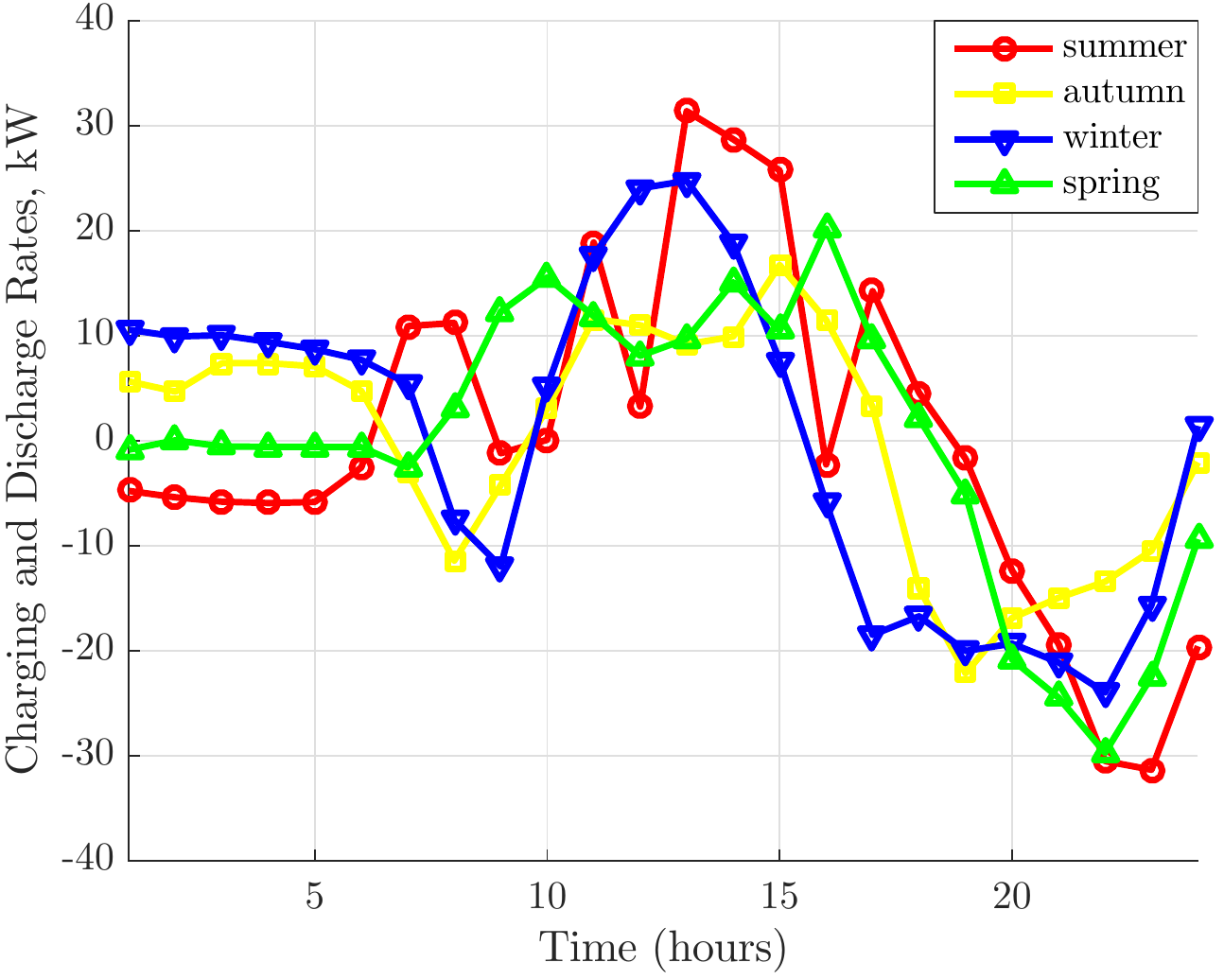}
	\caption{Charging and discharging rates on a typical day in each season}
	\label{fig2}
\end{figure}

\begin{figure}[] 
	\centering
	\includegraphics[width=8cm,keepaspectratio]{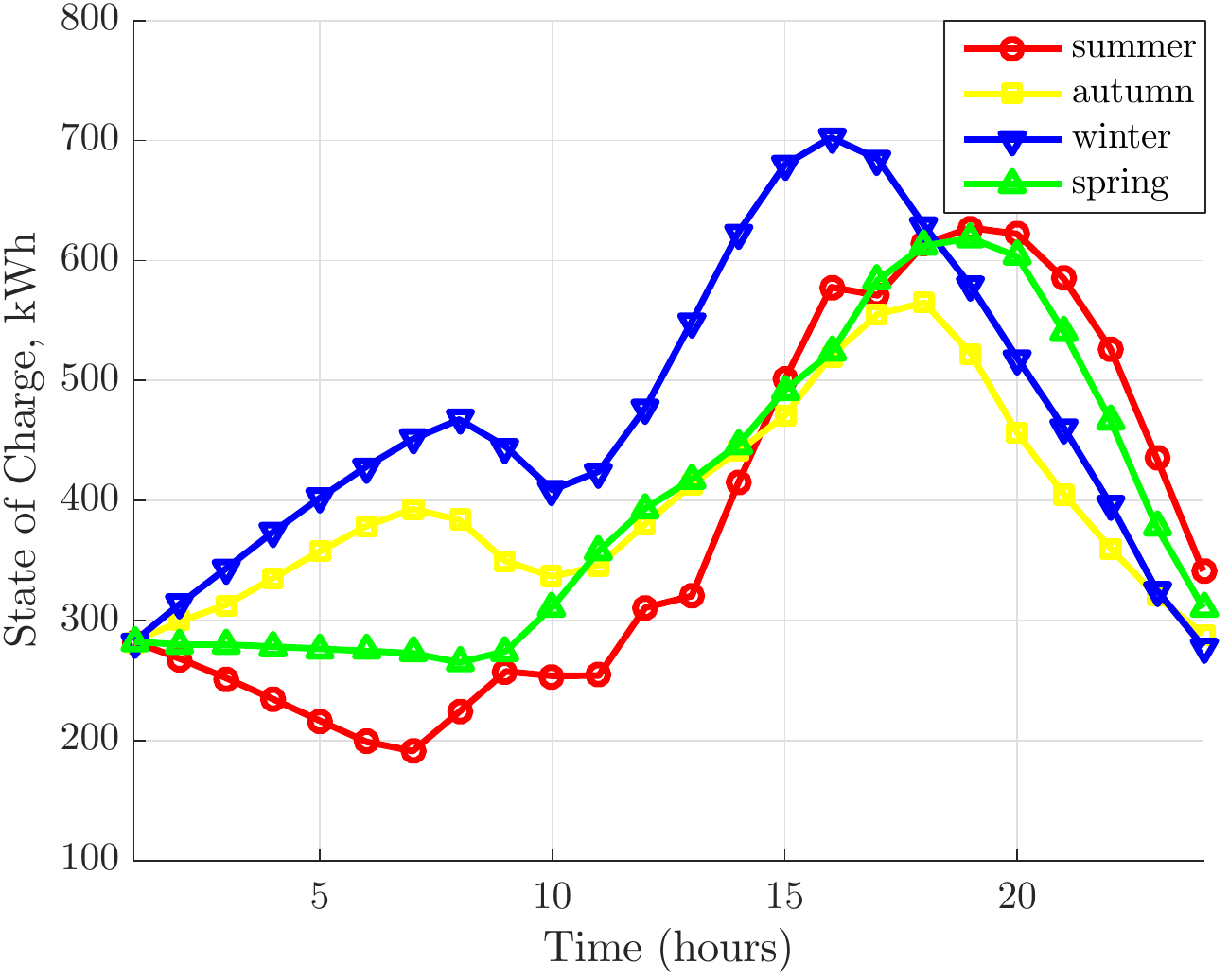}
	\caption{State of Charge on a typical day in each season}
	\label{fig3}
\end{figure}

Optimization-based simulations are performed for one year, and the outcomes suggest an installation at bus \SI{702} with the capacity being \num{942}\si{\kilo\watt\hour}. The initial state of charge for each cycle is \num{282}\si{\kilo\watt\hour}. The annual network energy loss in this case drops to \num{9410}\si{\kilo\watt\hour} which shows a \num{22.9}\si{\%} reduction compared with the base scenario. The detailed charging and discharging patterns of a sample day for each season are observed and presented in Fig.~\ref{fig2}. It can be seen that summer presents the highest charging rates around mid-day due to high PV penetration. Meanwhile, the discharging period comes at around 4pm in winter as night falls faster, while this time is delayed in summer as expected. Due to the long discharging period in winter, the ESS must charge in early mornings to maintain the state of charge on the pre-defined level at the beginning of each cycle. The corresponding state of charge is illustrated by Fig.~\ref{fig3}, from which we observe that the winter curve reaches its peak at \si{850\kilo\watt\hour} at around 5pm, earliest and highest among all, then drops rapidly due to the high energy consumption.

\subsubsection{Hosting Capacity}
The ESS with the same charging and discharging patterns is implemented in OpenDSS to quantify the hosting capacity in this scenario. Applying the same method as in the base scenario, an over voltage is discovered at bus \SI{2266} on a typical summer day at 1pm with PV systems at all loads being rated at \SI{2.71}{\kilo\watt}. This shows that the CESS is capable of increasing the hosting capacity of the network by \SI{27.8}{\%}.      

\subsubsection{Network Unbalance}
The maximum ${\mathrm{VUF}}$  in this scenario is improved from \SI{0.401}{\%} to \SI{0.381}{\%}, and the average ${\mathrm{VUF}}$ is decreased from \SI{0.275}{\%} to \SI{0.262}{\%}. This improvement can become more helpful when the ${\mathrm{VUF}}$ increases to high levels due to high PV penetration.     

\subsection{Distributed Energy Storage Systems}
In this subsection, the results of installing multiple DESSs, which have the same aggregated size as the CESS, in the network are detailed and compared with the performance indices and the first scenario. 

\subsubsection{Power Losses}
\begin{figure}[]
	\centering
	\includegraphics[width=8cm,keepaspectratio]{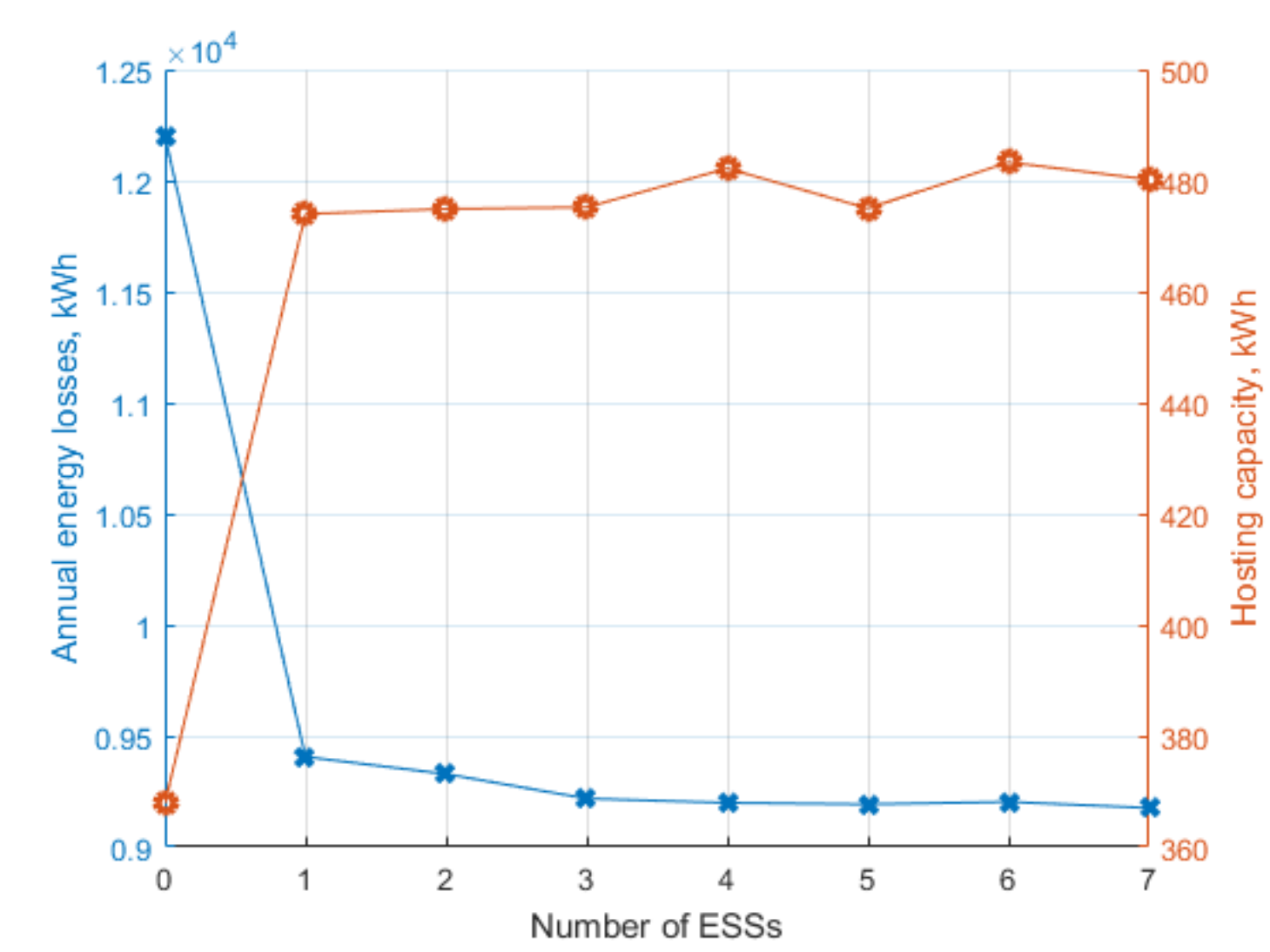}
	\caption{Annual energy losses (left) and hosting capacity (right) wrt number of ESSs}
 	\label{fig4}	
\end{figure}

Table~\ref{T1} and Fig.~\ref{fig4} indicate the improvements in annual energy losses and the hosting capacity with respect to different numbers of DESSs. An obvious decrease in energy losses is discovered as the CESS is implemented. However, as we continue to raise the number of ESS, the reduction immediately plateaus. When three ESSs are installed, energy loss is decreased only by \SI{190}{\kilo\watt\hour} to \SI{9220}{\kilo\watt\hour} from the previous scenario. After this, the reduction becomes subtle. This is because the aggregated size is fixed which is limiting the performance of the DESSs. Thus, larger numbers of ESSs require further increase in the aggregated capacity to become more effective. 

Greater number and size of DESSs potentially provide greater power loss reductions, however, these also lead to higher installation cost. Usually, if the installation cost is considerable and the utilities expect the ESS to be up-gradable for greater performance in the future, multiple DESSs can be applied. Further, the locations for DESSs are fairly uniformly distributed on the main branch of the feeder, which allows the DESSs to cover larger areas and become more effective. An example is shown in Fig.~\ref{fig1} where the optimal locations of three DESSs are indicated by green squares. 

\begin{table}[]
	\caption{Improvement in energy losses and the hosting capacity wrt different numbers of ESSs}
	\centering
	\begin{tabular}{lllll}
		\\[-1.8ex]\hline 
		\hline \\[-1.8ex] 
		No of   & Locations       	&Sizes \si{\kilo\watt\hour}       	    & Power loss    & Increase in  					\\
		ESSs 	& 					&	      								& reduction	    & hosting  				\\
				&					&										&				& capacity				\\
		\hline \\[-1.3ex] 				
		1		& 702				&942							& 22.9\%		& 27.8\%					\\
		\hline \\[-1.3ex] 			
		2       & 234, 1704        	&527, 414          				& 23.6\%      & 29.1\%                       \\
		\hline \\[-1.3ex] 
		3       & 153, 676, 1786    &322, 451        				& 24.5\%      & 29.2\%                       \\
				&					&169																				\\
		\hline \\[-1.3ex] 
		4       & 170, 652,		    &322, 427,    					& 24.6\%      & 31.1\%                       \\
				& 996, 1786			&114, 78.3																			 \\				
		\hline \\[-1.3ex] 
		5       & 153, 429, 728,	&179, 269,	   					& 24.7\%      & 29.2\%                           \\
				& 996, 1786			&312, 63.7,																			 \\
				&					&117																				 \\
		\hline \\[-1.3ex] 
		6		& 234, 652, 820,	&320, 346,   					& 24.8\%	   & 31.4\% 						 \\	
				& 1484, 1851, 2191	&114, 57.7,																		 \\
				&					&49.8, 54.9																		\\
	    \hline \\[-1.3ex] 
	    7		& 153, 429, 728,	&179, 269,			    		& 24.8\%	   & 30.6\% 						 \\	
				& 996, 1484, 1786,	&312, 39.3,																			 \\
				& 1924				&34.4, 17.4,																			 \\
				&					&89.7																					\\
		\hline \\[-1.3ex] 
	\end{tabular}
\label{T1}
\end{table}

\subsubsection{Hosting Capacity}
The results presented in Table~\ref{T1} and Fig.~\ref{fig4} show that there is an obvious increase, \SI{27.8}{\%}, in the hosting capacity after the CESS is installed. However, this rate reduces when the large ESS is divided into DESSs. Specifically, after two installations, the curve experiences minor fluctuations and the increase becomes subtle. The utilities must consider this with the trade-off between the ESS costs and performance introduced previously for deciding the most beneficial number of installations.  

\subsubsection{Network Unbalance}
Table~\ref{T2} shows the maximum and average ${\mathrm{VUFs}}$  with respect to different numbers of ESSs. The maximum ${\mathrm{VUF}}$, discovered at bus \SI{1940}, remains almost unchanged after replacing the CESS with multiple DESSs. Meanwhile, the average ${\mathrm{VUF}}$  drops from \SI{0.262}{\%} to \SI{0.261}{\%} if three installations are applied. Beyond this number, the average ${\mathrm{VUF}}$ does not change. The difference between the CESS and DESSs is almost negligible, and both scenarios present insignificant impacts on the network unbalance because this study adopts three-phase ESS which charge and discharge equal amount of power on each phase. We expect this result to be improved when single-phase ESS are used instead.

\begin{table}[] 
	\centering
	\caption{Max and average ${\mathrm{VUFs}}$  wrt numbers of ESSs}
	\begin{tabular}{lllll}
		\\[-1.8ex]\hline 
		\hline \\[-1.8ex]
		No of ESSs & Max ${\mathrm{VUF}}$  (\%) & Location (bus no.) & Average ${\mathrm{VUF}}$  (\%)   \\
		\hline \\[-1.3ex]
		0          & 0.401        & 1940               & 0.275              \\
		\hline \\[-1.3ex]
		1          & 0.381        & 1940               & 0.262              \\
		\hline \\[-1.3ex]
		2          & 0.382        & 1940               & 0.262              \\
		\hline \\[-1.3ex]
		3          & 0.381        & 1940               & 0.261              \\
		\hline \\[-1.3ex]
		4          & 0.381        & 1940               & 0.261              \\
		\hline \\[-1.3ex]
		5          & 0.381        & 1940               & 0.261              \\
		\hline \\[-1.3ex]
		6          & 0.381        & 1940               & 0.261              \\
		\hline \\[-1.3ex]
		7          & 0.381        & 1940               & 0.261              \\
		\hline \\[-1.3ex]
	\end{tabular}
	\label{T2}
\end{table}

\subsection{Comparisons of community and distributed ESSs}

Overall speaking, both CESS and DESSs can accomplish the tasks described in this study. Minor improvement can be made when attempting to decrease the ${\mathrm{VUF}}$ by replacing the CESS with multiple DESSs, while this improvement is larger regarding power losses and the hosting capacity. It is important to realize that the performance can be improved significantly when increasing the aggregated capacity of distributed ESSs. However, this is not the case for the CESS, for whom a further increase in size will result in more energy losses. Therefore, in general, DESSs can be more effective than CESS if the correct number of DESSs is placed at optimal locations with reasonable sizes, and consequently, they can make distribution upgrade deferral and increased PV generation a reality.  

\section{Conclusion}
This paper developed an MIQP model to find the optimal locations with their associated sizes for a CESS and multiple DESSs in an unbalanced LV network. A linearized approximation of branch flow equations is applied. The study analyzes, quantifies and compares the effectivenesses of minimizing power losses while increasing the hosting capacity between the two scenarios. The outcomes of simulations illustrated that both DESSs and CESS are effective in accomplishing these tasks, especially in reducing power losses and increasing the hosting capacity. When comparing the two ESS configurations, which are of the same aggregated size, against each other, the performance is marginally better for the DESSs. We expect that this can be improved significantly if further increase in the size of the DESSs is allowed.

\bibliographystyle{IEEEtran}
\bibliography{Ref}
%




\end{document}